\documentclass{ws-ijmpa}
\usepackage[super,compress]{cite}
\begin{document}
\def\mean#1{\left< #1 \right>}

\markboth{E.Conti, G.Sartori}
{On the coherent emission of radio frequency radiation}

\title{ON THE COHERENT EMISSION OF RADIO FREQUENCY RADIATION FROM HIGH ENERGY PARTICLE SHOWERS}

\author{ENRICO CONTI}
\address{INFN, Sezione di Padova, Via Marzolo 8 \\ I-35131 Padova, Italy
\\enrico.conti@pd.infn.it}

\author{GIORGIO SARTORI}
\address{Dipartimento di Fisica e Astronomia ``G.Galilei'', Universit\`a di Padova, Via Marzolo 8 \\ I-35131 Padova, Italy \\ giorgio.sartori@unipd.it}

\maketitle
\begin{abstract}

Extended Air Showers produced by cosmic rays impinging on the earth atmosphere irradiate radio frequency radiation through different mechanisms. Upon certain conditions, the emission has a coherent nature, with the consequence that the emitted power is not proportional to the energy of the primary cosmic rays, but to the energy squared. The effect was predicted in 1962 by Askaryan and it is nowadays experimentally well established and exploited for the detection of ultra high energy cosmic rays.

In this paper we discuss in details the conditions for coherence, which in literature have been too often taken for granted, and calculate them analytically, finding a formulation which comprehends both the coherent and the incoherent emissions. We apply the result to the Cherenkov effect, obtaining the same conclusions derived by Askaryan, and to the geosynchrotron radiation.

\end{abstract}

\ccode{PACS numbers:}

\section{Introduction}

The emission of radio frequency (RF) radiation from the Extended Air Showers (EAS) produced by Ultra High Energy Cosmic Rays (UHECR, i.e., particles with energy $E_p \gtrsim 10^{17}$eV) impinging on the earth atmosphere has been investigated since the 60s, after Askaryan \cite{Ask1962, Ask1965} proposed a coherent mechanism for the production of Cherenkov radiation at MHz frequencies (the so called Askaryan effect). The first detection of RF radiation from EAS was proved in 1965 by Jelley et al. \cite{Jelley1965}. The idea of Askaryan relies on the excess of negative charges in the shower, since positrons tend to disappear through their annihilation into photons, and further electrons are created via Compton scattering by photons. If the number of electrons and positrons is the same, emission would not occur because the electric field generated by opposite charges is equal (in module) and opposite (in sign). Upon certain conditions, the Cherenkov yield is not proportional to the number $N$ of charged particles in the shower, but to $(\eta N)^2$,  where $\eta$ is the fractional excess of charge. The quadratic dependence on $N$ (and therefore on $E_p$) makes the Askaryan effect attractive as mechanism for the detection of UHECRs.

In the following years, other mechanisms were proposed, which predict a coherent emission of RF radiation, and invoke the interaction of the shower electrons and positrons with the earth magnetic field. Negative and positive charges are separated by the magnetic field, creating an electric dipole. During the motion of the dipole, RF radiation is emitted in coherent way (geomagnetic effect). Alternatively, charged particles are bent because of the Lorentz force and emit synchrotron radiation (geosynchrotron effect). 

The bunch of particles in the EAS arriving to the detector can be schematically depicted as a dish, with thickness  $L_{dish}$ of a few meters and diameter depending of the energy of the primary. 
As will see in the followings, the requirement that the wavelength $\lambda$ is comparable with the dish thickness puts a frequency cutoff above which coherence is lost (and, consequently, the emission strongly hampered): $\nu \lesssim c/L_{dish} \sim 100$~MHz, where $c$ is the speed of light.  
The coherence condition explains why the experimental investigation is mostly concentrated in the frequency range $\lesssim 100$~MHz.

This picture is not complete. Only very recently, it has been shown, with Monte Carlo simulations \cite{deVries2011, AlvarezMuniz2012, Werner2012},  that geosynchrotron radiation can be coherent above $\sim$100~MHz under certain conditions: since the emission occurs in air, which has a (altitude dependent) refractive index $> 1$, a time compression of RF wavefront is produced, which causes the coherence to extend to higher frequency, even beyond the GHz. The compression takes place only inside a cone with aperture $\psi_c \approx 1^{\circ}$ equal to the Cherenkov angle, and angular spread $\Delta\psi \approx 0.1^{\circ}$. Outside this small angular region, coherence is lost and the RF output power falls dramatically. In this coherence regime, sometimes named ``geomagnetic Cherenkov radiation",  the emitted electric field amplitude $\mathbb{E}$ depends on the frequency and can be modelled, as shown by Refs. \refcite{AlvarezMuniz2012,ANITA2015}, as
 \begin{equation*}
\mathbb{E}(\nu)=\mathbb{E}(\nu_0)\exp\left(-\frac{\nu-\nu_0}{\nu_{\tau}}\right)
\end{equation*}
with $\nu_0= $ 300~MHz and $\nu_{\tau}\approx $ 500~MHz.
Here we will not deal with such particular mechanism.

The first experimental proof of the coherent Cherenkov emission in laboratory controlled conditions was performed in 2001 by directing picosecond pulses of GeV photons against a silica sand target \cite{Saltzberg2001}. Successively other tests were conducted using rock salt \cite{Gorham2005} or ice \cite{Gorham2007} as targets. Because of the higher density of those materials, the shower dimensions are different than in air, and Askaryan effect can extend beyond GHz frequencies.

The coherent RF emission by charged particles in the presence of a magnetic field has been measured in controlled conditions and successfully checked versus electrodynamics simulations in Ref. \refcite{Belov2016}.

Beside coherent emission, a RF incoherent emission mechanism exists, that is, bremsstralhung radiation, the emission of photons during the deflection of electrons and positrons under the coulomb field of the nuclei. The emission is always present but is weak since it scales linearly with $E_p$  and the cross section is low. 

In this paper we do not discuss the various  mechanisms of production of RF radiation and their relative importance. Our purpose is to explicitly show how and when coherent mechanisms appear, and to derive a general formula which involves the coherent and the incoherent regimes.

\section{Coherence}

We suppose that the EAS is composed by $N$ charged particles divided into $N_-$ electrons and $N_+$ positrons ($N = N_- + N_+$), and indicate with $\eta$ the charge excess: $\eta = (N_- - N_+)/N$ ($\eta \approx 0.1\div 0.25$).

Depending of the RF radiation mechanism, the phase of the electromagnetic wave from a single emitter (electron or positron) can change randomly with time or cannot. For bremsstrahlung, for example, the phase changes randomly and therefore the superposition of $N$ sources is incoherent and the emitted energy is proportional to $N$\footnotemark. For synchrotron, geomagnetic, and Cherenkov emission the phase of a single emitter does not change randomly with time. The superposition of $N$ emitters can be coherent or incoherent depending on the random relative position of the individual sources.

\footnotetext {Strictly speaking, bremsstrahlung emission can be coherent if the mean distance $\ell$ between two nearby emissions is much shorter than the wavelength $\lambda$. Since $\ell$ is of the order of $10^2$~m (Ref.~\refcite{Conti2014}), the condition on the frequency is $\nu \ll 3$~MHz.}

To simplify the calculations, we suppose that the electric field amplitude is unity and the fields are  polarized in the same direction.

For the case of  Cherenkov radiation, the electron and positron electric fields have opposite sign. Therefore the total electric field $\mathbb{E}$ (at the detector) is
\begin{equation}
 \label{eq:efield}
\mathbb{E} = \sum_{i=1}^{N_-}e^{i(\Phi_i-\omega t)} - \sum_{i=1}^{N_+}e^{i(\Theta_i-\omega t)} 
\end{equation}
where $\Phi_i$ and $\Theta_i$ are the phases of the RF wave emitted by the $i$-th electron and positron, respectively. For any $i$, $\Theta_i$ and $\Phi_i$ are independent random variables with the same distribution $f$, that is, 
$f(\Phi_1) = f(\Phi_2) = ...  = f(\Phi_{N_-})=f(\Phi) = f(\Theta_1) = f(\Theta_2)= ... = f(\Theta_{N_+})= f(\Theta)$. Therefore their mean values, for any $i$, are $\mean{\Phi_i}= \mean{\Phi} =\mean{\Theta_i}=\mean{\Theta}$.
The energy $W$ is 
\begin{eqnarray*}
&W = \mathbb{E}\mathbb{E}^* =  
\sum_{i,j=1}^{N_-}e^{i(\Phi_i-\Phi_j)} + \sum_{i,j=1}^{N_+}e^{i(\Theta_i-\Theta_j)} \\&- \sum_{i=1}^{N_+}e^{i(\Theta_i-\omega t)}\sum_{j=1}^{N_-}e^{-i(\Phi_j-\omega t)} 
- \sum_{i=1}^{N_-}e^{i(\Phi_i-\omega t)} \sum_{i=1}^{N_+}e^{-i(\Theta_i-\omega t)} 
\\
&= \sum_{i,j=1}^{N_-}e^{i(\Phi_i-\Phi_j)} + \sum_{i,j=1}^{N_+}e^{i(\Theta_i-\Theta_j)} -2\sum_{i=1}^{N_-}\sum_{j=1}^{N_+}\cos(\Phi_i-\Theta_j) \\&= W_1 + W_2 + W_{12}
\end{eqnarray*}
We are interested to the mean value $\mean{W} = \mean{W_1} + \mean{W_2} +\mean{W_{12}}$:
\begin{eqnarray*}
\mean{W_1}&=\mean{\sum_{i,j=1}^{N_-}e^{i(\Phi_i-\Phi_j)} } = N_- + \mean{\sum_{i\neq j}^{N_-}e^{i(\Phi_i-\Phi_j)} } \\&=
N_- + \mean{\sum_{i>j}^{N_-}(e^{i(\Phi_i-\Phi_j)} + e^{-i(\Phi_i-\Phi_j)}) } 
\\&= N_- + 2 \mean{\sum_{i>j}^{N_-}\cos(\Phi_i-\Phi_j)} = N_- + \sum_{i\neq j}^{N_-}\mean{\cos(\Phi_i-\Phi_j)}
\end{eqnarray*}
Since $\Phi_i$ is independent of $\Phi_j$ ($i\neq j$), $\mean{\cos(\Phi_i-\Phi_j)}$  =  $\mean{\cos\Phi_i\cos\Phi_j+\sin\Phi_i\sin\Phi_j}$ = $\mean{\cos\Phi_i}\mean{\cos\Phi_j} +  \mean{\sin\Phi_i}\mean{\sin\Phi_j}$ = $\mean{\cos\Phi_i}^2 + \mean{\sin\Phi_i}^2$.
It is reasonable to assume that the probability density $f(\Phi)$ is an even function between $-\Phi_0$ and 
$\Phi_0$, so $\mean{\sin\Phi_i} = 0$. In conclusion, 
\begin{eqnarray*}
\mean{W_1} = N_- +  \sum_{i\neq j}^{N_-}\mean{\cos(\Phi_i)}^2 = N_- + N_-(N_--1)\mean{\cos\Phi}^2
\end{eqnarray*}
Analogously 
\begin{eqnarray*}
\mean{W_2} &= N_+ +  \sum_{i\neq j}^{N_+}\mean{\cos(\Theta_i)}^2 = N_+ + N_+(N_+-1)\mean{\cos\Theta}^2 \\&= N_+ + N_+(N_+-1)\mean{\cos\Phi}^2
\end{eqnarray*}
\begin{eqnarray*}
\mean{W_{12}}&= -2\sum_{i=1}^{N_-}\sum_{j=1}^{N_+}\mean{\cos(\Phi_i-\Theta_j) }= -2N_+N_-\mean{\cos(\Phi - \Theta)} 
\\&= -2N_+N_-\mean{\cos\Theta}\mean{\cos\Phi} = -2N_+N_-\mean{\cos\Phi}^2 
\end{eqnarray*}
The energy $\mean{W}$ is therefore
\begin{equation}
    \label{eq:Wmedio}
\mean{W} = (N_- + N_+) +[(N_- - N_+)^2 - (N_- + N_+)]\mean{\cos\Phi}^2
\end{equation}

The factor which determines the presence of the coherence is precisely  $\mean{\cos\Phi}^2$. It can be easily seen in the case where the phases $\Phi$ are uniformly distributed between 
$-\Phi_0$ and $\Phi_0$:
\begin{equation*}
\mean{\cos\Phi} = \frac{1}{2\Phi_0}\int_{-\Phi_0}^{\Phi_0}\cos\Phi d\Phi =  \frac{\sin\Phi_0}{\Phi_0} 
\end{equation*}
Two extreme regimes can be identified:
\begin{enumerate}
\item $\Phi_0 \gg 1$: $\mean{\cos\Phi} = 0$, which is the condition for complete incoherence. The total energy is $\mean{W} = N_- + N_+ = N$. This is the case, for example, of the optical and UV Cherenkov light emission.
\item $\Phi_0 \to 0$: total coherence. $\mean{W} = (N_- - N_+)^2 = (\eta N)^2$, depends on the excess $\eta$ of negative charges in the shower.
\end{enumerate}

\begin{figure}[!t]
\centering
\includegraphics[scale=0.65]{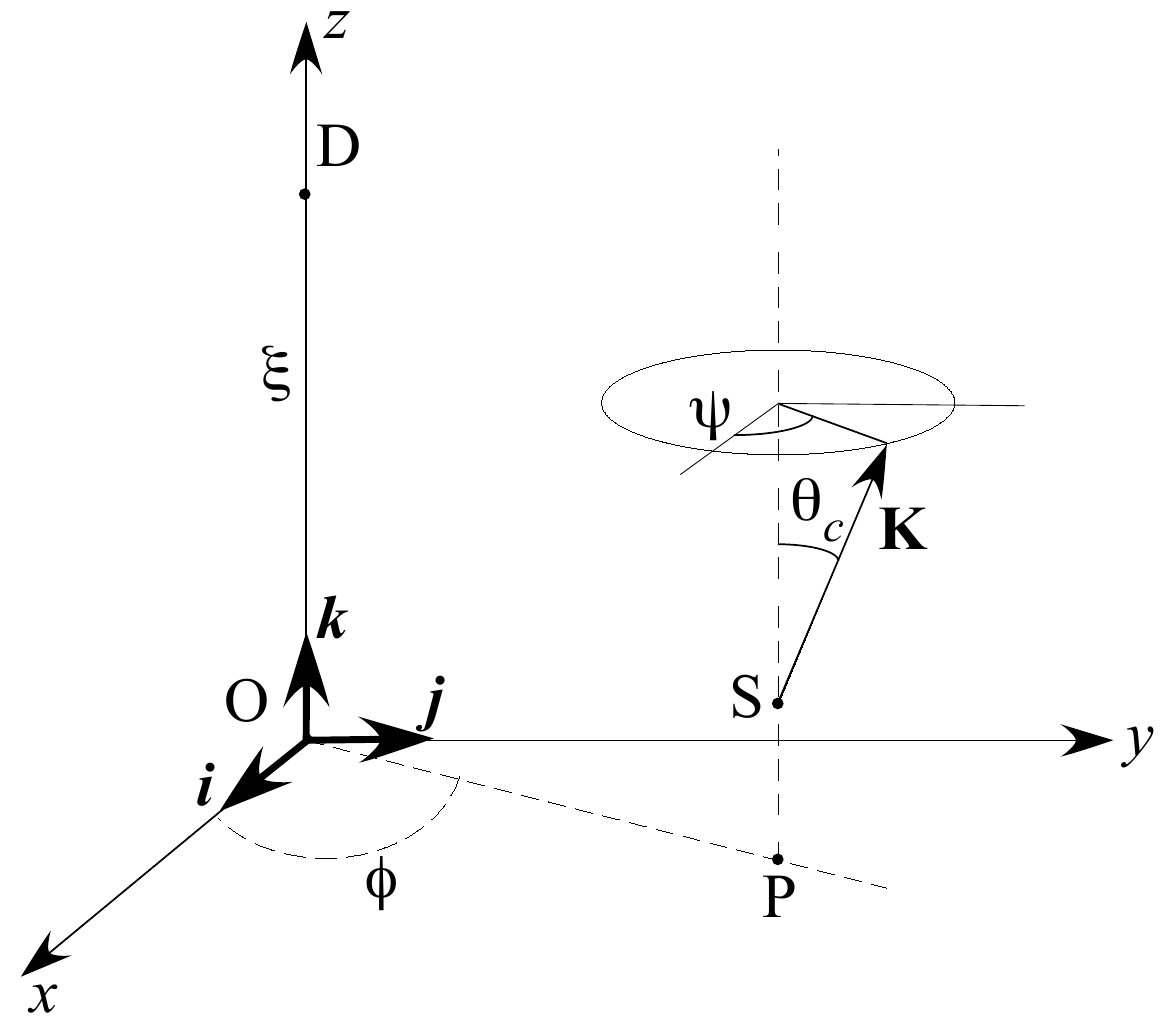}
\caption{
The geometry for the calculation of the coherent Cherenkov emission from an EAS. S is the emission point, D the detector position, and $\theta_c$ the Cherenkov angle.
}.
\label{fg:systemref}
\end{figure}

We treat now the generic case for the Cherenkov radiation where the phase $\Phi$ is not uniformely distributed. The results can be easily adapted to the geosynchronous and the geomagnetic radiation. With reference to Fig.\ref{fg:systemref}, we introduce a reference system with versors $\{\hat{i},\hat{j},\hat{k}\}$, where the EAS axis coincides the $z$ axis. $S$ is a single emitter with cylindrical coordinates $\{\rho,\phi,z\}$, $\theta_c$ is the Cherenkov cone aperture, $\psi$ the angular coordinate along the emission cone. The detector D is placed on the axis $z$ at the distance $\xi$: $\vec{OD} = \xi\hat{k}$. P is the projection of S on the x-y plane. Then: $\vec{r} \equiv \vec{SD}=\vec{OD}-\vec{OS}$, $\vec{OS}= \rho\cos\phi\hat{i} + \rho\sin\phi\hat{j} + z\hat{k}$. The wave vector $\vec{\mathbb{K}}$ is  $\vec{\mathbb{K}} =  {\mathbb{K}}(\sin\theta_c\sin\psi\hat{i} + \sin\theta_c\cos\psi\hat{j} + \cos\theta_c\hat{k})$ and the phase $\Phi$ of the electric field in the point D is 
\begin{equation*}
\Phi = \vec{\mathbb{K}}\cdot\vec{r} =   {\mathbb{K}}[-\rho\sin\theta_c\sin(\phi+\psi) + (\xi - z)]
\end{equation*}
The phase $\Phi$ depends also on $\xi$ (fixed value), but since in the calculation of the energy only the phase differences enter, the term with $\xi$ always eliminates, so we can put $\xi = 0$ to simplify the computation:
\begin{equation*}
\Phi = \vec{\mathbb{K}}\cdot\vec{r} =   -{\mathbb{K}}[\rho\sin\theta_c\sin(\phi+\psi) + z]
\end{equation*}
We schematize the shower as a flat cylinder, with thickness $L_{dish}$ and charge distribution $F(z)$  along the shower axis ($-{L_{dish}}/{2}\le z \le +{L_{dish}}/{2}$), and radial charge distribution $\sigma(r)$ such as $\int_0^\infty\sigma(r)rdr$ =1.
The angular variables $\phi$ and $\psi$ are distributed uniformily between 0 and $2\pi$.
Then
\begin{equation*}
\mean{\cos\Phi} = \int_0^{2\pi}\frac{d\phi}{2\pi}\int_0^{2\pi}\frac{d\psi}{2\pi}\int_{-L_{dish}/2}^{L_{dish}/2}dz \int_0^{\infty}d\rho F(z) \rho\sigma(\rho)\cos\Big[{\mathbb{K}}[\rho\sin\theta_c\sin(\phi+\psi) + z]\Big]
\end{equation*}
The computation is straightforward in the case $F(z) = {1}/{L_{dish}}$ (uniform distribution):
\begin{equation*}
\mean{\cos\Phi}=\frac{\sin (\mathbb{K} L_{dish}/2)}{\mathbb{K} L_{dish}/2}\int_0^{\infty}\rho\sigma(\rho)J_0(\mathbb{K}\rho\sin\theta_c)d\rho
\end{equation*}
where $J_0(x)$ is the Bessel function of the first kind. Calling:
\begin{equation}
 \label{eq:Az}
A_z(\mathbb{K},L_{dish})=\frac{\sin (\mathbb{K} L_{dish}/2)}{\mathbb{K} L_{dish}/2}
\end{equation}
and
\begin{equation}
 \label{eq:Ar}
A_r(\mathbb{K},\sigma) = \int_0^{\infty}\rho\sigma(\rho)J_0(\mathbb{K}\rho\sin\theta_c)d\rho
\end{equation}
then
\begin{equation}
 \label{eq:cosmedio}
\mean{\cos\Phi}= A_z(\mathbb{K},L_{dish})\cdot A_r(\mathbb{K},\sigma) \equiv A(\mathbb{K},L_{dish},\sigma)
\end{equation}

The same procedure can be applied to the geosynchrotron  emission. 
The electric field of the synchrotron radiation is \cite{Landau}:
\begin{equation*}
\vec{\mathbb{E}} = \frac{e}{4\pi\epsilon_0 c^2} \Bigg[\frac{(\hat{n}\times[(\hat{n}-\vec{\beta}) \times \vec{a}]}{r~(1-\vec{\beta}\cdot\hat{n})^3} \Bigg]_{retarded}
\end{equation*}
where $\hat{n}$ is the versor of the direction of observation, $\vec{a}=c\dot{\vec{\beta}}$ is the particle acceleration, and the electron electric charge $e$ must be considered with the sign. Since positrons and electrons have opposite acceleration under magnetic field, the product $e\vec{a}$ is the same for the two particles, so the electric fields sum up.
Therefore, for geosynchrotron radiation, the Eq.(\ref{eq:efield}) becomes
\begin{equation*}
\mathbb{E} = \sum_{i=1}^{N_-}e^{i(\Phi_i-\omega t)} + \sum_{i=1}^{N_+}e^{i(\Theta_i-\omega t)} 
\end{equation*}
 and the final result is:
\begin{equation}
 \label{eq:Wmedio2}
\mean{W} = (N_- + N_+) +[(N_- + N_+)^2 - (N_- + N_+)]\mean{\cos\Phi}^2 
\end{equation}
In case of total coherence, $\mean{W} = N^2$.

Eq. (\ref{eq:Wmedio}) and (\ref{eq:Wmedio2}) can be summarized stating that
the total power emitted by $N$ sources is obtained multiplying the power from a single source by a factor $M$:
\begin{equation}
    \label{eq:M}
M \approx N + (N_- \pm N_+)^2\cdot A(\nu)^2
\end{equation}
where $A(\nu) \equiv A(\mathbb{K},L_{dish},\sigma)$ represents the spatial coherence factor, function of the radiation frequency $\nu$ and of the shower shape and dimension, and the sign $\pm$ depends on the emission mechanism (-- for Cherenkov, + for synchrotron and geomagnetic).

The  factor $A(\nu)$ in Eq.(\ref{eq:M}) determines the coherent or incoherent regime. If 
 $A=0$, the incoherence is total, and $M = N$, i.e. the total emitted power is the some of the $N$ single powers. On the opposite, complete coherence occurs as $|A|  \rightarrow 1$, and $M = (\eta N)^2$ for Cherenkov emission, $M = N^2$ for geosynchrotron and geomagnetic emission, i.e., the power is proportional to the squared number of emitters.

The factorization ($\ref{eq:cosmedio}$) has the physical meaning that the longitudinal and radial developments of the showers are independent of each other, which is a reasonable approximation that can be extended to the general case.  $A_r$ ($|A_r| \leq 1$),   takes into account the radial emitter distribution $\sigma(r)$, in the plane perpendicular to the shower axis, while $A_z$  ($|A_z| \leq 1$)   takes into account the distribution along the longitudinal axis.  

As a check, note that when $\mathbb{K} \to \infty$, $A_z \to 0$ and $A_r \to 0$ (total spatial incoherence) and when $\mathbb{K} \to 0$, $|A_z| \to 1$ and  $|A_r| \to 1$ (total spatial coherence).

\section{Cherenkov effect in EAS}

As an example of the above calculations, we apply them to the case of EAS produced by cosmic rays in the atmosphere.

It is well known that a relativistic particle with velocity $\upsilon$ such that $\beta= \upsilon/c \geq 1/n_{air}$, where $n_{air}$ is the refractive index of air, radiates electromagnetic waves at the angle $\theta_c = \arccos(1/n_{air}\beta)$ at the energy rate (in SI units):
\begin{equation*}
\frac{d^2E}{dxd\lambda}=\frac{4\pi^2e^2}{\lambda^3}\left(1-\frac{1}{n_{air}^2\beta^2}\right)
\end{equation*}
Since $dx=cdt$,  $\beta\approx1$, and $n_{air}^2-1\approx2(n_{air}-1)$, for an EAS with $N$ secondary particles the radiated power $P$ in the bandwidth $\Delta\nu$ centred around the frequency $\nu$, neglecting coherence effets, is:
\begin{equation*}
P=4\pi\alpha h\cdot N \cdot(n_{air}-1)\nu\Delta\nu
\end{equation*}

\begin{figure}[t]
\centering
\includegraphics[scale=0.75]{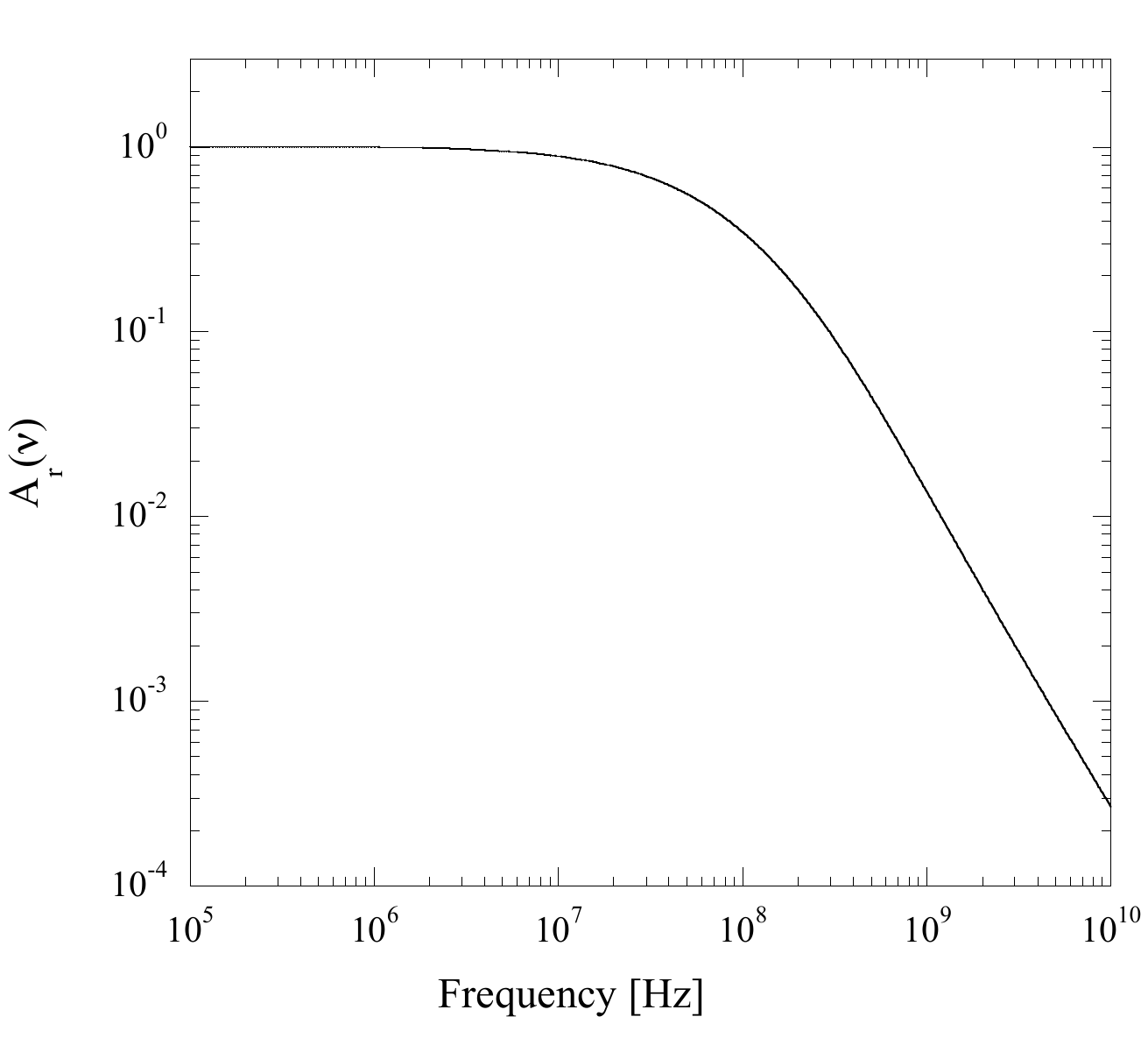}
\caption{
The function $A_r(\nu)$ for the coherent Cherenkov effect assuming the NKG radial charge distribution.
}
\label{fg:figNKG}
\end{figure}
\noindent
where $\alpha$ is the fine-structure constant and $h$ the Planck constant.

When coherence occurs, the emitted power is obtained from the above equation substituting the factor $N$ with $(\eta N)^2$ and taking into consideration the amplitude $A(\nu)$ (Eq.(\ref{eq:M})). Then
\begin{equation*}
P =~4\pi\alpha h\cdot (\eta N)^2\cdot  A_r^2 \cdot A_z^2\ \cdot(n_{air}-1)\nu\Delta\nu 
\end{equation*}
\noindent
To calculate $A_r(\nu,\sigma)$ from Eq.(\ref{eq:Ar}) we must know the radial charge distribution $\sigma$. Some examples can be found in the original work of Askaryan \cite{Ask1965}. Here we treat the case of the Nishima, Kamata, Greisen (NKG) distribution, which describes adequately the charge distribution in an EAS\cite{NKG1,NKG2}.
The normalized density function $\sigma(r)$  is:
\begin{equation*}
\sigma(r) =\frac{1}{2\pi R_M^2}\frac{\Gamma(4.5-s)}{\Gamma(s)\Gamma(4.5-2s)}\cdot\left(\frac{r}{R_M}\right)^{s-2}\left(1+\frac{r}{R_M}\right)^{s-4.5}
\end{equation*}
where  $\Gamma(t)$ is the factorial Gamma function,  $s=1.62$, and $R_M=28.1$ m.
$A_r(\nu)$ has been evaluated numerically and the result shown in Fig.\ref{fg:figNKG}. The frequency cutoff, where the power is reduced by a factor of 2, is $\nu \approx 30$ MHz.

Regarding the longitudinal coherence, in the simple case of an uniform charge distribution, we get, as shown before in Eq.(\ref{eq:Az}), a $(\sin x/x)$ function with the first zero occurring at $\nu_1=c/L_{dish}=100$ MHz for $L_{dish}=3$ m. Beyond that frequency, the RF emitted power scales as $1/\nu^2$. More realistic cases are discussed in Ref.~\refcite{HF2003}.

\section{Conclusions}
We have derived a general formula for the emission of radio frequency from particle showers, which 
describes also the transition from the incoherent regime to the coherent one. A coherence factor is introduced, which  depends on the spatial charge distribution of the radio frequency emitters.
The formula applies also to  dense materials, such as ice, for example, which has been proposed as target medium (precisely, Antartic ice sheet) the for the detection of very high energy neutrinos \cite{Frichter1996, RICE1997, ANITA2009}.

\end{document}